\documentclass[11pt]{article}
\usepackage{eurosym}
\usepackage{amsmath}
\usepackage{amsfonts}
\usepackage{amssymb}
\usepackage{graphicx}
\usepackage{epstopdf}
\usepackage{cite}
\usepackage[usenames]{color}
\usepackage{multirow}

\providecommand{\U}[1]{\protect\rule{.1in}{.1in}}
\newcommand{\ba}{\begin{array}}
\newcommand{\ea}{\end{array}}

\newcommand{\Dsl}[1] { \setbox0=\hbox{$#1$}     
\dimen0=\wd0   \setbox1=\hbox{/} \dimen1=\wd1  \ifdim\dimen0>\dimen1        
 \rlap{\hbox to \dimen0{\hfil/\hfil}}  #1 \else \rlap{\hbox to \dimen1{\hfil$#1$\hfil}}  /  \fi  }
\newcommand{\bea}{\begin{eqnarray}}
\newcommand{\eea}{\end{eqnarray}}

\newcommand {\nb}{\bar n}

\begin{document}

\title{ {\Large Baryon decays $J/\psi\rightarrow B\bar{B}$
within the QCD factorisation framework }}
\author{ Nikolay Kivel \\
\textit{\ Physik-Department, Technische Universit\"at M\"unchen,}\\
\textit{James-Franck-Str. 1, 85748 Garching, Germany } }
\maketitle

\begin{abstract}
We investigate  $J/\psi$  decays into octet baryon-antibaryons pairs. 
 The decay amplitudes are  computed within the collinear QCD factorisation framework. 
 The subleading amplitude, which describes the decay of longitudinally polarised charmonium,
  is computed  using  twist-4 three-quark distribution amplitudes.  The obtained results are
  used for a qualitative analysis of the experimental data.  It is found that  the polarisation 
  parameter $\alpha_B$  can be described  with an  accuracy  $10-30\%$, 
  which may indicate that the pQCD contribution dominates this observable.      
 \end{abstract}

\noindent

\vspace*{1cm}

\newpage

\section{Introduction}

\label{int}

An understanding of exclusive charmonia decays still remains challenging and
includes many open questions, see {\it e.g.} Refs.$\,$\cite{Brambilla:2004wf, Brambilla:2010cs}.
 A description of the underlying QCD dynamics for charmonia  is  complicated because the charm quark mass is not sufficiently large.
  On the other hand  this flaw   yields a possibility to measure observables,
which would be  much more difficult to access in case of heavier bottomonium.  This
open a window to study many interesting interplays of the long and short
distance QCD dynamics.  For instance, many observables associated with the
helicity flip amplitudes, which are strongly suppressed in the limit $
m_{Q}\rightarrow \infty $ ,  in case of charmonia decays can be accessed
and experimentally studied with a sufficiently high accuracy.  The decay $
J/\psi \rightarrow B\bar{B}$ into baryon-antibaryon pair  is the one
interesting example of such processes.  The final state baryons can be
easily detected  and existing  big  statistics of samples, which has been already collected at
BESIII allows one to measure  not only the branching ratio but also the
polarisation parameter $\alpha _{B}$, which describes the angular behaviour of
the cross section 
\begin{equation}
\frac{dN}{d\cos \theta }=A(1+\alpha _{B}\cos ^{2}\theta ),
\end{equation}%
where $\theta $ is the angle between the baryon or antibaryon direction and
the lepton beam, $A$ is an overall normalisation. The value of $\alpha _{B}$
is sensitive to the value of the helicity flip amplitude, which describes
the decay of the longitudinally polarised $S$-wave charmonia.  In the naive limit $m_{Q}\rightarrow \infty $  this quantity is given by $%
\alpha _{B}\rightarrow 1+O(m_{B}^{2}/m_{Q}^{2})$ \cite{Brodsky:1981kj}.  
The experimental results obtained for various baryon channels indicate
that  values of $\alpha _{B}$ definitely differ from one, see Refs.$\,$\cite{BES:2008hwe,BESIII:2012ion, BESIII:2016ssr, BESIII:2017kqw,BESIII:2020fqg}. In Tab.$\,$\ref{dataBB} we
summarise the existing data for the octet baryons.
 \begin{table}[th]
\centering
\caption{ The experimental data for decays $J/\psi\to B\bar{B}$. The values of the branching ratios are taken from PDG \cite{Zyla:2020zbs}. } 
\begin{tabular}{|c|c|c|c|}
\hline
$B$ & Br$[J/\psi \rightarrow B\bar{B}]\times 10^{3}$ & $\alpha _{B}$ & $%
Q_{B}=\frac{\text{Br}[\psi (2S)\rightarrow B\bar{B}]}{\text{Br}[J/\psi
\rightarrow B\bar{B}]}\times 100$ \\ \hline
$p$ & $2.12(3)$ & $0.59(1)$ \cite{BESIII:2012ion} & $13.86(3)$ \\ \hline
$n$ & $2.1(2)$ & $0.50(4) $ \cite{BESIII:2012ion}& $14.6(1)$ \\ \hline
$\Lambda $ & $1.89(9)$& $0.47(3)$\cite{BESIII:2017kqw}  & $20.4(1)$ \\ \hline
$\Sigma ^{0}$ & $1.17(3)$  & $-0.45(2)$\cite{BESIII:2017kqw} & $21.0(3)$ \\ \hline
$\Sigma ^{+}$ & $1.5(3)$& $-0.51(2)$  \cite{BESIII:2020fqg} & $7.2(5)$ \\ \hline
$\Xi ^{+}$ & $0.97(8)$ & $0.58(4)$ \cite{BESIII:2016ssr} & $26.7(5)$ \\ \hline
\end{tabular}%
\label{dataBB}
\end{table}

The last column in this table shows the ratio $Q_{B}$, which can be related
to the similar leptonic ratio 
\begin{equation}
Q_{l}=\frac{\text{Br}[\psi (2S)\rightarrow l\bar{l}]}{\text{Br}[J/\psi
\rightarrow l\bar{l}]}\times 100=13.21.
\end{equation}%
QCD factorisation predicts that in the limit of large mass $m_{Q}\rightarrow
\infty $ 
\begin{equation}
Q_{l}=Q_{B}+\mathcal{O}(v^{2})+\mathcal{O}(\Lambda ^{2}/m_{Q}^{2}),
\end{equation}%
where $v$ is the heavy quark velocity in the quarkonium rest frame, $v^{2}\ll
1$ and $\Lambda $ is a typical hadronic scale.  Therefore if the
leading-oder contribution in the expansion with respect $1/m_{Q}$ works
sufficiently well, one finds%
\begin{equation}
Q_{B}\simeq Q_{l}\text{.}
\end{equation}
It turns out that this approximate equation is not well satisfied for many decay channels.
For instance, for many mesonic decays $J/\psi \rightarrow MM$ this ratio is
strongly violated, see {\it e.g.}  Ref.~\cite{Brambilla:2004wf}. From the Tab.$\,$\ref{dataBB}  one can see that this
criteria sufficiently well works for nucleon states but for  other octet
baryons  $Q_B$ is about factor 2 differ from $Q_l$.  However this violation effect  is not too big. For example, for some
 mesonic decays  this difference is about an order of  magnitude.  Therefore one can expect that the leading-order description for
the baryonic decays can provide sufficiently large or perhaps even the
dominant effect.

The first realistic estimate of the value of $\alpha _{N}$  was considered
in Ref.~\cite{Claudson:1981fj}, where it is suggested to neglect the subleading  helicity flip
amplitude but keep  finite the  ratio $m_{N}^{2}/M_{\psi }^{2}$.  Then one finds
\begin{equation}
\alpha _{N}\simeq \frac{1-4m_{N}^{2}/M_{\psi }^{2}}{1+4m_{N}^{2}/M_{\psi}^{2}}\simeq 0.46,
\end{equation}
 which is a better estimate  comparing with the naive limit $m_{Q}\rightarrow \infty $.  In Ref.~\cite{Carimalo:1985mw}
  $\alpha _{N}$ is estimated using an idea that  nucleon is the
non-relativistic bound state of three quarks and each quark carries
approximately $1/3$ of the nucleon momentum. The result  of this
consideration yields $\alpha _{N} \simeq 0.66$. 

A more sophisticated idea  has been used in Ref.~\cite{Murgia:1994dh}, where the QCD
factorisation is used in order to compute the decay amplitudes.  In this framework the $c\bar{c}$-pair annihilates
at short distances into three hard gluons, which create light quark-antiquark
pairs describing the long-distance collinear overlap with outgoing  nucleon and antinucleon states. The
non-perturbative physics is encoded by  well defined matrix elements, which
are closely associated with hadronic wave functions. Such framework allows
one to build a systematic description performing the expansion with respect to small velocity $v$ and
small ratio $\Lambda /m_{Q}$.  The contribution of the helicity flip amplitude is subleading
and it is suppressed by extra power $\Lambda ^{2}/m_{Q}^{2}$
because the corresponding hard subprocess with massless quarks is sensitive
to the orbital angular momentum of the light quarks, which provides  an additional  power
suppression.  In Ref.$\,$\cite{Murgia:1994dh} the helicity flip amplitude is
obtained by introduction of  the effective light quark masses, which are taken to be $ m_{q}\simeq
x_{i}m_{N}$, where $x_{i}$ are the longitudinal momentum fractions
satisfying $x_{1}+x_{2}+x_{3}=1$. All amplitudes, which are calculated in this model,  are
sensitive to  the leading-twist light-cone  distribution amplitude (DA) 
$\varphi _{3}(x_{1},x_{2},x_{3})$, which describes the sharing of the
nucleon longitudinal momentum between the quarks at zero transverse
separation. Such  calculation of  the helicity flip amplitude  can only be considered as a certain
phenomenological model.

In Ref.~\cite{Kivel:2019wjh} the helicity flip amplitude for the first time has been computed within
the more systematic framework: using the three-quark higher twist DAs, which
are related to the higher Fock components of the nucleon wave function. Such DAs can also  be interpreted as 
a  $P$-wave configurations of the  three collinear constituent quarks.
It turns out  that the subleading helicity flip amplitude  can be computed within the same collinear factorisation 
framework as the leading-order amplitude.   The obtained result has been used for the analysis of the proton-antiproton data. 
The numerical estimate obtained in Ref.~\cite{Kivel:2019wjh} is  $\alpha _{p}\simeq 0.71$,
which allows one to conclude that the factorisation description works
sufficiently well in this case.  Therefore charmonium decay data may also provide an interesting  information about the baryon twist-4 DAs.

Quite different idea has been developed in Refs.~\cite{Zhu:2015bha, Alekseev:2018qjg, BaldiniFerroli:2019abd}.  In these works the effective hadronic Lagrangian density has been constructed using the flavour $SU(3)$ symmetry arguments.  The unknown vertex couplings have been fitted from the data.  One of the main output of this analysis is the estimate of the relative phase between the hadronic and electromagnetic amplitudes, which are defined through $c\bar{c}\to ggg$ and  $c\bar{c}\to \gamma^*$ subprocesses, respectively. In Ref.~\cite{BaldiniFerroli:2019abd} this phase  is found to be  relatively closer to $\pi/2$ than to $0$ or $\pi$. Qualitatively this does not agree  with the factorisation picture, which predicts that the hadronic amplitudes are real.  Basing on this observation  it is  concluded  that this disagreement is perhaps an indication that  pQCD  can not provide a sufficiently good description of $J/\psi$ decays into baryon-antibaryon pairs.

In the present  work we are going to extend the analysis of the Ref.~\cite{Kivel:2019wjh} to  baryon-antibaryon states from the baryon octet.  The  results
for the hard kernels obtained in Ref.~\cite{Kivel:2019wjh} also allows one to calculate the amplitudes for  other 
baryons. Therefore for such analysis one only needs  the information about
the twist-3 and twist-4 DAs of the  baryons.  Some moments of these DAs
have been studied recently on the lattice in Ref.~\cite{Bali:2019ecy} and these results can
be used in order to constrain the non-perturbative input.  

The paper is organised as follows. In Sec.~\ref{BDA}  we discuss the baryon
DAs and describe the models, which are used in our calculations. The 
analytical expressions for the amplitudes and observables are discussed in the Sec.~\ref{decampl}.
 The discussion of the numerical results are presented in Sec.~\ref{phen}.  In
Sec.~\ref{conc} we provide the conclusions. In Appendix we provide  the useful details  about the baryon
DAs, which are used in our calculations.

\section{Baryon DAs}

\label{BDA}

The long distance dynamics associated with the QCD hadronisation into
outgoing hadrons is described by the matrix elements, which are parametrised
in terms of the light-cone distribution amplitudes (DAs). The structure and
the properties of the higher twist octet baryon DAs have already been studied in Refs.~\cite{Braun:2008ia, Anikin:2013aka,Wein:2015oqa,Anikin:2015qos}. 
Here we briefly discuss  the required twist-3 and twist-4 DAs
and construct the phenomenological models.

In the following we  assume that the baryon momentum $k$ is directed along $z$-axis and
can be expanded as 
\begin{equation}
 k=k_{-}\frac{{n}}{2}+k_{+}\frac{\bar {n}}{2},\ \ k_{+}\gg 
k_{-}=\frac{m_{B}^{2}}{k_{+}}, ~~ k^{2}=m_{B}^{2}.
\end{equation}%
Here $n,\bar{n}$ are the auxiliary light-cone vectors $n^{2}=\bar{n}^{2}=0$, 
$(n\bar{n})=2$ and 
\begin{equation}
k_{+}=(kn),\quad k_{-}=(k\bar{n}).
\end{equation}%
The baryon spinors $N(k,s)$ satisfy Dirac equations\footnote{%
We do not introduce for spinor the subscript indicating the baryon $B$,
assuming that this will be clear from the context.} 
\begin{equation*}
(\Dsl{k}-m_{B})N(k,s)=0,
\end{equation*}%
and normalised as $\bar{N}(k,s)N(k,s^{\prime })=2m_{B}\delta _{ss^{\prime }}$%
. It is also convenient to introduce the large and small components $N_{\bar{%
n}}$ and $N_{{n}}$, respectively: 
\begin{equation*}
N_{\bar{n}}=\frac{\setbox0=\hbox{$\bar n$}\dimen0=\wd0\setbox1=\hbox{/}\dimen%
1=\wd1\ifdim\dimen0>\dimen1\rlap{\hbox to \dimen0{\hfil/\hfil}}\bar{n}\else%
\rlap{\hbox to \dimen1{\hfil$\bar n$\hfil}}/\fi\setbox0=\hbox{$n$}\dimen0=\wd%
0\setbox1=\hbox{/}\dimen1=\wd1\ifdim\dimen0>\dimen1%
\rlap{\hbox to
\dimen0{\hfil/\hfil}}n\else\rlap{\hbox to \dimen1{\hfil$n$\hfil}}/\fi}{4}%
N(k,s),~\ ~N_{{n}}=\frac{\setbox0=\hbox{$n$}\dimen0=\wd0\setbox1=\hbox{/}%
\dimen1=\wd1\ifdim\dimen0>\dimen1\rlap{\hbox to \dimen0{\hfil/\hfil}}n\else%
\rlap{\hbox to \dimen1{\hfil$n$\hfil}}/\fi\setbox0=\hbox{$\bar n$}\dimen0=\wd%
0\setbox1=\hbox{/}\dimen1=\wd1\ifdim\dimen0>\dimen1%
\rlap{\hbox to
\dimen0{\hfil/\hfil}}\bar{n}\else\rlap{\hbox to \dimen1{\hfil$\bar n$\hfil}}/%
\fi}{4}N(k,s)=\frac{m_{B}}{k_{+}}\frac{\setbox0=\hbox{$n$}\dimen0=\wd0\setbox%
1=\hbox{/}\dimen1=\wd1\ifdim\dimen0>\dimen1%
\rlap{\hbox to
\dimen0{\hfil/\hfil}}n\else\rlap{\hbox to \dimen1{\hfil$n$\hfil}}/\fi}{2}N_{%
\bar{n}}.~
\end{equation*}%
The similar relations also hold for the antibaryon spinors.

In this paper we only consider the contributions of the three quark
operators. The contributions of the quark-gluon operators corresponds to the
moments with higher conformal spin and will be neglected \cite{Braun:2008ia, Anikin:2013aka}. The relevant
three-quark light-cone operators are constructed from the QCD quark fields $%
q={u,d,s}$ and light-cone the Wilson lines 
\begin{equation}
W_{\nb}[x_{-},z_{-}]=\text{P}\exp \left\{ ig\int_{(z_{-}-x_{-})/2}^{0}ds\
A_{+}(x_{-}n/2+sn)\right\} .
\end{equation}%
The light-cone three-quark operator can be written as 
\begin{equation}
\mathcal{O}_{\alpha _{1}\alpha _{2}\alpha
_{3}}(x_{-},y_{-},z_{-})=\varepsilon ^{ijk}\ q_{\alpha _{1}}^{i^{\prime
}}(x_{-})W_{\nb}[x_{-},v_{-}]_{i^{\prime }i}\ q_{\alpha _{2}}^{j^{\prime
}}(y_{-})W_{\nb}[y_{-},v_{-}]_{j^{\prime }j}\ q_{\alpha _{3}}^{k^{\prime
}}(z_{-})W_{\nb}[z_{-},v_{-}]_{k^{\prime }k},
\end{equation}%
where we used short notation $q(x_{-})\equiv q(x_{-}n/2)$. The set of
indices $\{ijk\}$ stands for the colour, the indices \ $\alpha _{i}$ refer
to the Dirac structure.  Following to Ref.~\cite{Wein:2015oqa} we assume the following
flavour content of the operators $\langle 0|q_{\alpha _{1}}q_{\alpha
_{2}}q_{\alpha _{3}}|B\rangle \equiv \langle 0|q_{1}q_{2}q_{3}|B\rangle $: 
\begin{eqnarray}
\langle 0|u_{1}u_{2}d_{3}|p\rangle ,\, \langle 0|u_{1}d_{2}s_{3}|\Lambda\rangle ,
\label{qqq1}
\\
 \langle 0|u_{1}d_{2}s_{3}|\Sigma ^{0}\rangle,\,
  \langle0|u_{1}u_{2}s_{3}|\Sigma ^{+}\rangle , 
  \label{qqq2}
\\
\langle 0|s_{1}s_{2}u_{3}|\Xi^{0}\rangle,\, \langle 0|s_{1}s_{2}d_{3}|\Xi ^{-}\rangle .
\label{qqq3}
\end{eqnarray}
 For simplicity
this will not be shown explicitly. For brevity  we also simplify the
notation for the Dirac indices assuming $\{\alpha _{1},\alpha _{2},\alpha
_{3}\}\equiv \{1,2,3\}$.

The universal light-cone matrix elements can be defined as  \cite{Braun:2008ia}%
\begin{eqnarray}  
&&
\left\langle 0\right\vert \mathcal{O}_{123}(x_{-},y_{-},z_{-})\left\vert B(k)\right\rangle  =
\nonumber\\[2mm]   &&
\frac{1}{4}\left[\Dsl{k} C\right] _{12}\left[ \gamma_{5}N_{\bar{n}}\right] _{3}\text{FT}\left[ V_{1}^{B}\right] 
+\frac{1}{4}\left[\Dsl{k} \gamma _{5}C\right] _{12} \left[ N_{\bar{n}}\right] _{3}\text{FT}\left[ A_{1}^{B}\right]
 +\frac{1}{4}\left[ \Dsl{k}\gamma _{\bot }^{\alpha }C\right] _{12} \left[\gamma _{\bot }^{\alpha }\gamma _{5}N_{\bar{n}}\right] _{3}~\text{FT}\left[ T_{1}^{B}\right]
\nonumber\\ [2mm]  &&
+\frac{m_{B}}{8k_{+}}\left[ \Dsl{k}C\right] _{12}~\left[ \gamma _{5} \Dsl{n}
N_{\bar{n}}\right] _{3}\text{FT}\left[ V_{2}^{B}\right] +\frac{m_{B}}{8k_{+}}%
\left[ \Dsl{k} \gamma _{5}C\right] _{12}~\left[ \Dsl{n}N_{\bar{n}}\right] _{3}\text{FT}\left[A_{2}^{B}\right]
\nonumber \\[2mm] &&
 +\frac{m_{B}}{8}\left[ \gamma _{\bot }^{\alpha }C\right] _{12}~\left[
\gamma _{\alpha \bot }\gamma _{5}N_{\bar{n}}\right] _{3}~\text{FT}\left[
V_{3}^{B}\right] +\frac{m_{B}}{8}\left[ \gamma _{\bot }^{\alpha }\gamma _{5}C%
\right] _{12}~\left[ \gamma _{\alpha \bot }N_{\bar{n}}\right] _{3}~\text{FT}%
\left[ A_{3}^{B}\right]
\nonumber \\[2mm] &&
+\frac{m_{B}}{4}~\left[ C\right] _{12}\left[ \gamma _{5}N_{\bar{n}}\right]
_{3}~\text{FT}\left[ S_{1}^{B}\right] +\frac{m_{B}}{4}~\left[ \gamma _{5}C%
\right] _{12}\left[ N_{\bar{n}}\right] _{3}~\text{FT}\left[ P_{1}^{B}\right]
\nonumber \\[2mm] &&
+\frac{m_{B}}{8k_{+}}\left[
\Dsl{k}\gamma _{\bot }^{\alpha }C\right] _{12}\left[
\Dsl{n}\gamma _{\bot }^{\alpha }\gamma _{5}N_{\bar{n}}\right] _{3}~\text{FT}\left[
T_{2}^{B}\right]
+\frac{m_{B}}{4k_{+}}\left[ i\sigma _{kn}C\right] _{12}\left[ \gamma _{5}
N_{\bar{n}}\right] _{3}\text{FT}\left[ T_{3}^{B}\right] 
\nonumber \\[2mm] &&
+\frac{m_{B}}{8}\left[
\sigma _{\alpha \beta }C\right] _{12}\left[ \sigma _{\bot \bot }^{\alpha
\beta }\gamma _{5}N_{\bar{n}}\right] _{3}~\text{FT}\left[ T_{7}^{B}\right] .
\label{meL6}
\end{eqnarray}%
where we always assume that 
\begin{equation*}
k\simeq k_{+}\frac{\bar n}{2},\ \sigma _{kn}=\sigma _{\alpha \beta }k^{\alpha
}n^{\beta }\ .
\end{equation*}%
The symbol ``$\bot$'' denotes the transverse projections with respect to the light-like vectors:
$n_{\alpha }\gamma _{\bot }^{\alpha }=\bar{n}_{\alpha }\gamma _{\bot}^{\alpha }=0$.
The symbol ``FT" denotes the Fourier transformation
\begin{equation}
\text{FT}[f]=\int
Du_{i}~e^{-i(k_{1}x)-i(k_{2}y)-i(k_{3}z)}f(u_{1},u_{2},u_{3}),~\   \label{FT}
\end{equation}%
with the measure $Du_{i}=du_{1}du_{2}du_{3}\delta (1-u_{1}-u_{2}-u_{3})$,
the quark momenta in (\ref{FT}) are defined as 
\begin{equation}
k_{i}=u_{i}k_{+}\frac{\bar{n}}{2}.
\end{equation}%
The defined in Eq.(\ref{meL6}) DAs  are symmetric/antisymmetric
with respect to interchange $u_{1}\leftrightarrow u_{2}$%
\begin{equation}
F_{i}^{B}(x_{2},x_{1},x_{3})=-(-1)_{B}F_{i}^{B}(x_{1},x_{2},x_{3}),\ \text{%
for }\ \ F=\{S,P,A\},  \label{AsDA}
\end{equation}%
\begin{equation}
F_{i}^{B}(x_{2},x_{1},x_{3})=+(-1)_{B}F_{i}^{B}(x_{1},x_{2},x_{3}),\ \text{%
for }\ \ F=\{V,T\},  \label{SymDA}
\end{equation}%
with%
\begin{equation}
(-1)_{B}=\left\{ 
\begin{array}{c}
+1\ \ \ B\neq \Lambda \\ 
-1\ \ \ B=\Lambda%
\end{array}%
\right. .\text{ }
\end{equation}

The minimal basic set of baryon light-cone DAs  can be defined in terms of
these DAs as \cite{Wein:2015oqa, Anikin:2015qos} ( below we accept the short notation $ (x_{1},x_{2},x_{3})\equiv (x_{123})$)%
\begin{equation}
f_{B}\varphi _{3\pm }^{B}(x_{123})=\frac{c_{B}^{\pm }}{2}\left\{ \left(
V_{1}-A_{1}\right) ^{B}(x_{123})\pm \left( V_{1}-A_{1}\right)
^{B}(x_{321})\right\} ,  \label{phi3Bpm}
\end{equation}%
\begin{equation}
f_{B}\Pi _{3}^{B}(x_{123})=c_{B}^{-}(-1)_{B}T_{1}^{B}(x_{132}),
\end{equation}%
\begin{equation}
\Phi _{4\pm }^{B}(x_{123})=c_{B}^{\pm }\left\{ \left( V_{2}-A_{2}\right)
^{B}(x_{123})\pm (-1)_{B}\left( V_{3}-A_{3}\right) ^{B}(x_{231})\right\} ,
\label{phi4pm}
\end{equation}%
\begin{equation}
\Xi _{4\pm }^{B}(x_{123})=(-1)_{B}3c_{B}^{\pm }\left[ ~\left(
T_{3}-T_{7}+S_{1}+P_{1}\right) ^{B}(x_{123})\pm \left(
T_{3}-T_{7}+S_{1}+P_{1}\right) ^{B}(x_{132})\right] ,  \label{Ksi4pm}
\end{equation}%
\begin{equation}
\Pi _{4}^{B}(x_{123})=c_{B}^{-}\left( T_{3}+T_{7}+S_{1}-P_{1}\right)
^{B}(x_{312}),  \label{Pi4}
\end{equation}%
\begin{equation}
\Upsilon _{4}^{B}(x_{123})=6c_{B}^{-}T_{2}^{B}(x_{321}),  \label{Ups4}
\end{equation}%
where  
\begin{equation}
c_{B}^{+}=\left\{ 
\begin{array}{c}
1,~B\neq \Lambda  \\ 
\sqrt{\frac{2}{3}},B=\Lambda 
\end{array}%
\right. ,~~\ c_{B}^{-}=\left\{ 
\begin{array}{c}
1,~B\neq \Lambda  \\ 
-\sqrt{6},B=\Lambda 
\end{array}%
\right. ,
\end{equation}%
and $f_{B}$ is the normalisation coupling. The coefficients $c_{B}^{\pm }$
are chosen in such way that these DAs satisfy the simple  relations in the 
$SU(3)$ flavour symmetry limit \cite{Wein:2015oqa}.

For the nucleon state, the basic set of DAs can be further simplified by  the
isospin relationships, namely,  it is sufficient to determine%
\begin{equation}
f_{N~}\varphi _{3}(x_{123})=\phi _{3+}^{N}(x_{123})+\phi
_{3-}^{N}(x_{123})=\left( V_{1}-A_{1}\right) ^{N}(x_{123}),  \label{phi3}
\end{equation}%
\begin{equation}
\frac{1}{2}\Phi _{4}(x_{123})=\frac{1}{2}\left( \Phi _{4+}^{N}+\Phi
_{4-}^{N}\right) (x_{123})=\left( V_{2}-A_{2}\right) ^{N}(x_{123}),
\label{Phi4}
\end{equation}%
\begin{equation}
\frac{1}{2}\Psi _{4}(x_{123})=\frac{1}{2}\left( \Phi _{4+}^{N}-\Phi
_{4-}^{N}\right) (x_{312})=\left( V_{3}-A_{3}\right) ^{N}(x_{231}),
\label{Psi4}
\end{equation}%
\begin{equation}
\frac{\lambda _{2}}{6}\Xi _{4}(x_{123})=\left( \Xi _{4+}^{N}+\Xi
_{4-}^{N}\right) (x_{123})=\left( T_{3}-T_{7}+S_{1}+P_{1}\right)
^{N}(x_{123}).  \label{Ksi4}
\end{equation}%
The nucleon DAs $\varphi _{3},$ $\Phi _{4},\Psi _{4}$ and $\Xi _{4}$ can be
defined\footnote{%
We do not write the superscript ``N" for these DAs following to the notations
accepted in the literature.}  in terms of the matrix elements of the appropriately projected light-cone operators 
as in  Refs.$\,$\cite{Braun:2008ia, Anikin:2013aka}. Let us notice that some definitions of the DAs
in Ref.$\,$\cite{Braun:2000kw} differ from ones in used Ref.$\,$\cite{Anikin:2013aka}, which are  also used in this
paper.

The twist-4 DAs in Eqs (\ref{phi4pm})-(\ref{Ups4}) can be decomposed into
contributions associated with the contributions of the operators with the \textit{geometrical} twist-3 and twist-4.
Such decomposition is often referred in the literature as  Wandzura-Wilczek one.  In this paper we write such decompositions in
the following form\footnote{%
Let us notice that this choice is  different from one  accepted in
Ref~\cite{Wein:2015oqa}.}%
\begin{equation}
\Phi _{4\pm }^{B}(x_{123})=f_{B}\Phi _{4\pm }^{B(3)}(x_{123})+\lambda
_{1}^{B}\ \bar{\Phi}_{4\pm }^{B}(x_{123}),  
\label{WWPhi}
\end{equation}%
\begin{equation}
\Pi _{4}^{B}(x_{123})=f_{\bot }^{B}\ \Pi _{4}^{B(3)}(x_{123})+\lambda
_{1}^{B}\bar{\Pi}_{4}^{B}(x_{123}),\ \ B\neq \Lambda ,  \label{WWPiB}
\end{equation}%
\begin{equation}
\Pi _{4}^{\Lambda }(x_{123})=f_{\Lambda }\Pi _{4}^{\Lambda
(3)}(x_{123})+\lambda _{\bot }^{\Lambda }\bar{\Pi}_{4}^{\Lambda }(x_{123}).
\label{WWPiL}
\end{equation}%
\begin{equation}
\Xi _{4\pm }^{B}(x_{123})=\lambda _{2}^{B}\ \bar{\Xi}_{4\pm }^{B}(x_{123}),
\label{WWKsi}
\end{equation}%
\begin{equation}
\Upsilon _{4}^{B}(x_{123})=\lambda _{2}^{B}\bar{\Upsilon}_{4}^{B}(x_{123}).
\label{WWUps}
\end{equation}%
The functions in the rhs of Eqs.(\ref{WWPhi})-(\ref{WWUps}) are dimensionless
while  the couplings $f_{B},\ f_{\bot }^{B},\ \lambda _{1,2}^{B}$ and $%
\lambda _{\bot }^{\Lambda }$ have dimension  GeV$^{2}$.  The functions
with the superscript ``$(3)$"  describe  the contribution of geometrical
twist-3, i.e. these functions are completely defined in terms of the moments
of twist-3 DAs. The explicit expressions can be found in Refs.\cite{Braun:2008ia,Anikin:2013aka}. 

  The equations (\ref{phi3Bpm})-(\ref{Ups4}) can be easily rewritten using the symmetry properties (\ref{AsDA})-(\ref{SymDA})  in order to get the
expressions for the DAs $\{V_{i},A_{i},T_{i},S_{i},P_{i}\}$ in terms of the
basic functions $\{\varphi _{3\pm }^{B},\Pi _{3}^{B},\Phi _{4\pm }^{B},\Xi
_{4\pm }^{B},\Pi _{4}^{B},\Upsilon _{4}^{B}\}$. Therefore the non-perturbative input
associated with the final baryon and antibaryon states can be unambiguously described  in terms of
the basic DAs.  The construction of the  models for these functions are based on the truncate d
conformal expansions, see {\it e.g.} Refs.~\cite{Braun:2008ia, Anikin:2013aka}.  
 
For the nucleon DAs we take the model set ``ABO1" suggested in Ref.~\cite{Anikin:2013aka}.
These models have been designed in order  to describe nucleon
electromagnetic FFs within the light-cone sum rules. The twist-3 DA reads 
\begin{eqnarray}
\varphi _{3}(x_{123})&=&120x_{1}x_{2}x_{3}~\left( 1+\phi _{10}\mathcal{P}%
_{10}(x_{123})+\phi _{11}\mathcal{P}_{11}(x_{123})
\nonumber\right.  \\ 
&&+\quad \left. \phi _{20}\mathcal{P}%
_{20}(x_{123})+\phi _{21}\mathcal{P}_{21}(x_{i})+\phi _{22}\mathcal{P}%
_{22}(x_{123})\right),  \label{phi3N}
\end{eqnarray}%
where the orthogonal polynomials $\mathcal{P}_{kl}(x_{i})$ are given in Appendix. 
The moments $\phi _{kl}$ depends on the scale $\mu$ and they are multiplicatively
renormalisable, see the details in Appendix.   The chiral-even twist-4 DAs (\ref{Phi4}) and (\ref{Psi4}) can be represented as 
\begin{equation}
\Phi _{4}(x_{123})=f_{N}\Phi _{4}^{(3)}(x_{123})+\lambda _{1}\bar{\Phi}%
_{4}(x_{123}),  \label{Phi4N}
\end{equation}%
\begin{equation}
\Psi _{4}(x_{123})=f_{N}\Psi _{4}^{(3)}(x_{123})-\lambda _{1}\bar{\Psi}%
_{4}(x_{123}),  \label{Psi4N}
\end{equation}%
where the functions $\Phi _{4}^{(3)}$ and $\Psi _{4}^{(3)}$ are defined in
terms of twist-3 moments $\phi _{kl}$.  The models for the genuine twist-4 DAs reads
\begin{equation}
\bar{\Phi}_{4}(x_{123})=24x_{1}x_{2}\left( 1+\eta _{10}\mathcal{R}%
_{10}(x_{312})-\eta _{11}\mathcal{R}_{11}(x_{312})\right) ,  \label{PhiN4bar}
\end{equation}%
\begin{equation}
\bar{\Psi}_{4}(x_{123})=24x_{1}x_{3}\left( 1+\eta _{10}\mathcal{R}%
_{10}(x_{231})+\eta _{11}\mathcal{R}_{11}(x_{231})\right) .  \label{PsiN4bar}
\end{equation}%
The moments $\eta _{ik}$ are multiplicatively renormalisable and do not mix
with the moments from the quark-gluon DAs.  The explicit  expressions for
the polynomials $\mathcal{R}_{1i}$ and $\Phi _{4}^{(3)}$ and $\Psi _{4}^{(3)}$ are 
also given in  Appendix. The chiral-odd DA  $\Xi^B _{4}$ and $\Upsilon^B_4$  (\ref{Ksi4}) 
do not contribute at this order because the corresponding  hard kernels are trivial \cite{Kivel:2019wjh}.  

For other baryons we consider  more simple models. The   twist-3 DAs are described as 
\begin{equation}
\varphi _{3+}^{B}(x_{i})=120x_{1}x_{2}x_{3}~\left( 1+\phi _{11}^{B}\mathcal{P%
}_{11}(x_{123})\right) ,\ \ \ \varphi
_{3-}^{B}(x_{i})=120x_{1}x_{2}x_{3}~\phi _{10}^{B}\mathcal{P}_{10}(x_{123}).
\label{phiB3pm}
\end{equation}%
\begin{equation}
\Pi _{3}^{\Lambda }(x_{123})=120x_{1}x_{2}x_{3}~\pi _{10}^{\Lambda }\mathcal{P}_{10}(x_{123}),
\label{PiL3}
\end{equation}%
\begin{equation}
\Pi _{3}^{B}(x_{123})=120x_{1}x_{2}x_{3}(1+~\pi _{11}^{B}\mathcal{P}%
_{11}(x_{123})),\ \  B=\Sigma ,\Xi . 
 \label{PiB3}
\end{equation}%
The twist-3 moments $\{ \phi _{10}^{B},\phi _{11}^{B}, \pi _{\Lambda}^{B},\pi _{11}^{B} \}$ in Eqs.(\ref{phiB3pm})-(\ref{PiB3})  also enter in the expressions for the twist-4 DAs in Eqs.(\ref{WWPhi})-(\ref{WWPiL}), 
more details can be found in Appendix.  For the  genuine twist-4 parts  we use the following models
\begin{equation}
\bar{\Phi}_{4+}^{B}(x_{123})=24~x_{1}x_{2}\left( -\eta _{11}^{B}\right) 
\mathcal{R}_{11}(x_{312}),  \label{PhiB4pl}
\end{equation}%
\begin{equation}
\bar{\Phi}_{4-}^{B}(x_{123})=~24~x_{1}x_{2}\left( 1+\eta _{10}^{B}\mathcal{R}%
_{10}(x_{312})\right) ,
\end{equation}%
\ \ 
\begin{equation}
\bar\Pi _{4}^{\Lambda }(x_{123})=24x_{1}x_{2}~\left( 1+\zeta _{10}^{\Lambda }%
\mathcal{R}_{10}(x_{132})\right) ,
\end{equation}%
\begin{equation}
\ \bar{\Pi}_{4}^{B}(x_{i})=24~x_{1}x_{2}\left( -\zeta _{11}^{B}\right) 
\mathcal{R}_{11}(x_{312}),\ B=\Sigma ,\Xi .  \label{PiB4}
\end{equation}%

  The described DAs in the limit
of exact flavour symmetry $SU(3)$ ($m_{u}=m_{d}=m_{s}$)  must satisfy \cite{Wein:2015oqa}
\begin{equation}
\Phi _{i+}^{N\ast }=\Phi _{i+}^{\Sigma \ast }=\Phi _{i+}^{\Xi \ast }=\Phi
_{i+}^{\Lambda \ast }=\Pi _{i}^{N\ast }=\Pi _{i}^{B\ast },\ \ i=3,4
\label{SU3Phi+}
\end{equation}%
\begin{equation}
\Phi _{i-}^{N\ast }=\Phi _{i-}^{\Sigma \ast }=\Phi _{i-}^{\Xi \ast }=\Phi
_{i-}^{\Lambda \ast }=\Pi _{i}^{\Lambda \ast },\ \ \ i=3,4.  \label{SU3Phi-}
\end{equation}%
In terms of \ the moments this gives 
\begin{equation}
f_{N}^{\ast }=f_{\Sigma }^{\ast }=f_{\Xi }^{\ast }=f_{\bot }^{\Sigma \ast
}=f_{\bot }^{\Xi \ast },
\end{equation}%
\begin{equation}
\lambda _{1}^{\ast }=\lambda _{1}^{\Lambda \ast }=\lambda _{1}^{\Sigma \ast
}=\lambda _{1}^{\Xi \ast }=\lambda _{\bot }^{\Lambda \ast },
\end{equation}%
\begin{equation}
\phi _{10}^{\ast }=\phi _{10}^{\Lambda \ast }=\phi _{10}^{\Sigma \ast }=\phi
_{10}^{\Xi \ast }=\pi _{10}^{\Lambda \ast },
\end{equation}%
\begin{equation}
\eta _{10}^{\ast }=\eta _{10}^{\Lambda \ast }=\eta _{10}^{B\ast }=\zeta
_{10}^{\Lambda \ast },
\end{equation}%
\begin{equation}
\eta _{11}^{\ast }=\eta _{11}^{\Lambda \ast }=\eta _{11}^{\Sigma \ast }=\eta
_{11}^{\Xi \ast }=\zeta _{11}^{\Sigma \ast }=\zeta _{11}^{\Xi \ast }.
\end{equation}%
These relations are useful in the phenomenological analysis. The
numerical values of the required moments will be discussed later. 

  The analytical result for the helicity flip amplitude, which involves
the twist-4 DAs, can be described in terms of the so-called auxiliary DAs,
which are the linear combinations of the basic DAs. Such combinations  naturally appear in the calculation of hard kernels and allows one to derive
a compact analytical expressions for the amplitudes. 
These auxiliary  DAs can be defined in terms of the matrix elements of the twist-4 operators
with the  transverse derivative, see the details in Ref.~\cite{Kivel:2019wjh}.  The
corresponding operators are defined using the gauge invariant collinear
fields, which have  definite scaling behaviour within the effective field
theory framework 
\begin{equation}
\chi (x_{-})=W_{\bar{n}}^{\dag }(x_{-})\frac{\bar{n}n}{4}q(x_{-}),\ \ \ \ \
W_{\bar{n}}(x_{-})=\text{P}\exp \left\{ ig\int_{-\infty }^{0}ds\
A_{+}(x_{-}n/2+sn)\right\} .
\end{equation}%
Then  the auxiliary DAs can be introduced  through the following light-cone matrix
elements  
\begin{equation}
~\left\langle 0\right\vert \left[ i\partial _{\bot }^{\alpha }\chi (x_{-})%
\right] C\setbox0=\hbox{$n$}\dimen0=\wd0\setbox1=\hbox{/}\dimen1=\wd1%
\ifdim\dimen0>\dimen1\rlap{\hbox to \dimen0{\hfil/\hfil}}n\else\rlap{\hbox
to \dimen1{\hfil$n$\hfil}}/\fi\chi (y_{-})\chi _{3}(z_{-})\left\vert
B(k)\right\rangle =k_{+}m_{B}\left[ \gamma _{\bot }^{\alpha }\gamma _{5}N_{%
\bar{n}}\right] _{3}\text{FT}\left[ \mathcal{V}_{1}^{B}\right] ,
\label{calV1}
\end{equation}%
\begin{equation}
\left\langle 0\right\vert ~\chi (x_{-})C\setbox0=\hbox{$n$}\dimen0=\wd0%
\setbox1=\hbox{/}\dimen1=\wd1\ifdim\dimen0>\dimen1\rlap{\hbox to
\dimen0{\hfil/\hfil}}n\else\rlap{\hbox to \dimen1{\hfil$n$\hfil}}/\fi\left[
i\partial _{\bot }^{\alpha }\chi (y_{-})\right] \chi _{3}(z_{-})\left\vert
B(k)\right\rangle =k_{+}m_{B}~\left[ \gamma _{\bot }^{\alpha }\gamma _{5}N_{%
\bar{n}}\right] _{3}\text{FT}\left[ \mathcal{V}_{2}^{B}\right] .
\end{equation}%
\begin{equation}
\left\langle 0\right\vert \left[ i\partial _{\bot }^{\alpha }\chi (x_{-})%
\right] C\setbox0=\hbox{$n$}\dimen0=\wd0\setbox1=\hbox{/}\dimen1=\wd1\ifdim%
\dimen0>\dimen1\rlap{\hbox to \dimen0{\hfil/\hfil}}n\else\rlap{\hbox to
\dimen1{\hfil$n$\hfil}}/\fi\gamma _{5}\chi (y_{-})\chi _{3}(z_{-})\left\vert
B(k)\right\rangle =k_{+}m_{B}\left[ ~\gamma _{\bot }^{\alpha }N_{\bar{n}}%
\right] _{3}\text{FT}\left[ \mathcal{A}_{1}^{B}\right] ,
\end{equation}%
\begin{equation}
\left\langle 0\right\vert \chi (x_{-})C\setbox0=\hbox{$n$}\dimen0=\wd0\setbox%
1=\hbox{/}\dimen1=\wd1\ifdim\dimen0>\dimen1\rlap{\hbox to
\dimen0{\hfil/\hfil}}n\else\rlap{\hbox to \dimen1{\hfil$n$\hfil}}/\fi\gamma
_{5}\left[ i\partial _{\bot }^{\alpha }\chi (y_{-})\right] \chi
_{3}(z_{-})\left\vert B(k)\right\rangle =k_{+}m_{B}\left[ \gamma _{\bot
}^{\alpha }N_{\bar{n}}\right] _{3}\text{FT}\left[ \mathcal{A}_{2}^{B}\right]
.
\end{equation}%
\begin{align}
\left\langle 0\right\vert \left[ i\partial _{\bot }^{\alpha }\chi (x_{-})%
\right] C\setbox0=\hbox{$n$}\dimen0=\wd0\setbox1=\hbox{/}\dimen1=\wd1\ifdim%
\dimen0>\dimen1\rlap{\hbox to \dimen0{\hfil/\hfil}}n\else\rlap{\hbox to
\dimen1{\hfil$n$\hfil}}/\fi\gamma _{\bot }^{\nu }\chi (y_{-})\chi
_{3}(z_{-})\left\vert B(k)\right\rangle & =~k_{+}m_{B}~g_{\bot }^{\alpha \nu
}\left[ \gamma _{5}N_{\bar{n}}\right] _{3}\text{FT}\left[ \mathcal{T}%
_{21}^{B}\right]  \\
& +k_{+}m_{B}\left[ i\sigma _{\bot \bot }^{\nu \alpha }\gamma _{5}N_{\bar{n}}%
\right] _{3}\text{FT}\left[ \mathcal{T}_{41}^{B}\right] ,
\end{align}%
\begin{align}
\left\langle 0\right\vert \chi (x_{-})C\setbox0=\hbox{$n$}\dimen0=\wd0\setbox%
1=\hbox{/}\dimen1=\wd1\ifdim\dimen0>\dimen1\rlap{\hbox to
\dimen0{\hfil/\hfil}}n\else\rlap{\hbox to \dimen1{\hfil$n$\hfil}}/\fi\gamma
_{\bot }^{\nu }\left[ i\partial _{\bot }^{\alpha }\chi (y_{-})\right] \chi
_{3}(z_{-})\left\vert B(k)\right\rangle & =k_{+}m_{B}~g_{\bot }^{\alpha \nu }%
\left[ \gamma _{5}N_{\bar{n}}\right] _{3}\text{FT}\left[ \mathcal{T}_{22}^{B}%
\right]   \notag \\
& +k_{+}m_{B}\left[ i\sigma _{\bot \bot }^{\nu \alpha }\gamma _{5}N_{\bar{n}}%
\right] _{3}\text{FT}\left[ \mathcal{T}_{42}^{B}\right] ,  \label{def:calT2i}
\end{align}%
where the colour and the flavour structure of the operators are the same as
described before. The DAs in the\textit{\ rhs} of these equations can be
expressed in terms of  DAs $\{V_{i},A_{i},T_{i},S_{i},P_{i}\}^{B}$ using the
Lorentz symmetry and QCD equations of motion.  This gives \cite{Anikin:2013aka,Kivel:2019wjh}%
\begin{eqnarray}
4\mathcal{V}_{i}^{B}(x_{123})=x_{3}\left( V_{2}+A_{2}\right)
^{B}(x_{123})+(-1)^{i}\left[ (x_{1}-x_{2})V_{3}^{B}(x_{123})+\bar{x}%
_{3}A_{3}^{B}(x_{123})\right] 
\nonumber \\
+\frac{m_{3}}{m_{B}}\left(
V_{1}^{B}(x_{123})+(-1)^iA_{1}^{B}(x_{123})\right) +(-1)^{i+1}2\frac{m_{1}-m_{2}}{m_{B}}T_{1}^{B}(x_{123}),
\label{calVi}
\end{eqnarray}%
\begin{eqnarray}
4\mathcal{A}_{i}^{B}(x_{123})=-x_{3}\left( V_{2}+A_{2}\right)
^{B}(x_{123})+(-1)^{i}\left[ (x_{1}-x_{2})A_{3}^{B}(x_{123})+\bar{x}
_{3}V_{3}^{B}(x_{123})\right]
\nonumber \\
+\frac{m_{3}}{m_{B}}\left(
A_{1}^{B}(x_{123})+(-1)^iV_{1}^{B}(x_{123})\right)+(-1)^{i+1}2\frac{m_{1}+m_{2}}{m_{B}}T_{1}^{B}(x_{123}),
\end{eqnarray}
\begin{eqnarray}
\left( \mathcal{T}_{21}+\mathcal{T}_{41}\right) ^{B}(x_{123}) &=&\frac{x_{1}%
}{2}\left( T_{3}-T_{7}+S_{1}+P_{1}\right) ^{B}(x_{123}) -\frac{1}{2}\frac{m_{1}}{m_{B}}\left(V_{1}^{B}+A_{1}^{B}\right) ,\  
\\
\ \ \ \left( \mathcal{T}_{22}+\mathcal{T}_{42}\right) ^{B}(x_{123}) &=&\frac{x_{2}}{2}\left( T_{3}-T_{7}-S_{1}-P_{1}\right) ^{B}(x_{123})
-\frac{1}{2}\frac{m_{2}}{m_{B}}\left(V_{1}^{B}-A_{1}^{B}\right),
\end{eqnarray}%
\begin{eqnarray}
\left( \mathcal{T}_{21}-\mathcal{T}_{41}\right) ^{B}(x_{123}) &=&\frac{x_{1}%
}{2}\left( T_{3}+T_{7}+S_{1}-P_{1}\right) ^{B}(x_{123})+\frac{1}{2}\frac{m_{1}}{m_{B}}\left(A_{1}^{B}-V_{1}^{B}\right),\ \ \ 
 \\
\ \left( \mathcal{T}_{22}-\mathcal{T}_{42}\right) ^{B}(x_{123}) &=&
\frac{x_{2}}{2}\left( T_{3}+T_{7}-S_{1}+P_{1}\right) ^{B}(x_{123}) -\frac{1}{2}\frac{m_{2}}{m_{B}}\left(V_{1}^{B}+A_{1}^{B}\right),
\label{calT2i}
\end{eqnarray}%
where the quark masses $m_i$ correspond to the quarks $q_1q_2q_3$ in the matrix elements in Eqs.(\ref{qqq1})-(\ref{qqq3}).  
The contributions with the quark masses appears after application QCD equations for the  quark fields, see more details in Ref. \cite{Kivel:2019wjh}.
In what follow we assume $m_u=m_d\simeq 0$ and $m_s\neq 0$.  Therefore the terms with $m_s$ represent the part of the $SU(3)$-breaking corrections. 
Because the auxiliary DAs are uniformly defined  for all baryons, the description of decay amplitudes for different baryon
states uses  the same hard kernels, which are obtained for the nucleon case in Ref.~\cite{Kivel:2019wjh}.

\section{Decay amplitudes}
\label{decampl}

The decay amplitude $J/\psi (P)\rightarrow B(k)\bar{B}(k^{\prime })$ is
defined as 
\begin{equation}
M=\left( \epsilon _{\psi }\right) _{\mu }\ \bar{N}(k)\left\{ \gamma ^{\mu }\
A_{1}^{B}+(k^{\prime }+k)_{\nu }\frac{i\sigma ^{\mu \nu }}{2m_{B}}%
A_{2}^{B}\right\} V(k^{\prime })~,  \label{def:M}
\end{equation}%
where $\bar N$ and $V$ \ denote the baryon and antibaryon spinors, respectively.
The charmonium polarisation vector $\epsilon _{\psi }^{\mu }\equiv \epsilon
_{\psi }^{\mu }(P,\lambda )$ satisfies 
\begin{equation}
\sum_{\lambda }\epsilon _{\psi }^{\mu }(P,\lambda )\epsilon _{\psi }^{\nu
}(P,\lambda )=-g^{\mu \nu }+\frac{P^{\mu }P^{\nu }}{M_{\psi }^{2}},
\end{equation}%
where $P^{2}=M_{\psi }^{2}$. The observables can be conveniently described
in terms of  linear combinations 
\begin{equation}
\mathcal{G}_{M}^{B}=A_{1}^{B}+A_{2}^{B},\ \ \ \mathcal{G}_{E}^{B}=A_{1}^{B}+%
\frac{M_{\psi }^{2}}{4m_{B}^{2}}A_{2}^{B},
\end{equation}%
which are similar to the Sachs redefinition for the electromagnetic FFs.  These
amplitudes can be computed within the standard factorisation framework. The resulting 
expressions can be written as 
\begin{equation}
\mathcal{G}_{M/E}^{B}=\frac{f_{\psi }}{~m_{c}^{2}}\frac{~f_{B}^{~2}}{%
m_{c}^{4}}~(\pi \alpha _{s})^{3}~\frac{10}{81}\ (1+\delta _{\Lambda B})\
J_{M/E}^{B},  
\label{calGEMB}
\end{equation}%
where $f_{\psi }$ is the charmonium coupling defined below in Eq.(\ref%
{me:fpsi}), $\alpha _{s}$ is the QCD coupling and $\delta _{\Lambda B}=1\ \ $%
if $B=\Lambda $ and $\ \delta _{\Lambda B}=0\ \ $if $B\neq \Lambda .$ The leading-order
dimensionless convolution integrals read%
\begin{eqnarray}
 J_{M}^{B}=\frac{1}{4f_{B}^{2}}\int \frac{Dx_{i}}{x_{1}x_{2}x_{3}}\int 
\frac{Dy_{i}}{y_{1}y_{2}y_{3}}
&&\left\{
 \left( V_{1}-A_{1}\right)^{B}(x_{123})\left( V_{1}-A_{1}\right) ^{B}(y_{123})\frac{x_{1}y_{3}}{D_{1}D_{3}} 
 \right.
\nonumber \\ 
&&\left.+T_{1}^{B}(y_{123})T_{1}^{B}(x_{123})~\frac{2x_{1}y_{2}}
{D_{1}D_{2}} \right\} , 
 \label{JBM}
\end{eqnarray}%
where%
\begin{equation}
D_{i}=x_{i}(1-y_{i})+(1-x_{i})y_{i}.
\end{equation}%
This result has been obtained already long time ago, see {\it e.g.} Refs.~\cite{Chernyak:1987nv}. \ 

The second amplitude $\mathcal{G}_{E}^{B}$ has been  computed in
Ref.~\cite{Kivel:2019wjh} for the nucleon case. This result can be easily generalised to
other baryons taking into acount the universality of the definitions of the
baryon DAs \ (\ref{calV1})-(\ref{calT2i}).  Corresponding integrals read 
\begin{eqnarray}
 J_{E}^{B} &=&\frac{2}{f_{B}^{2}}\int
Dy_{i}~\frac{1}{y_{1}y_{2}y_{3}}\int Dx_{i}\frac{1}{x_{1}x_{2}x_{3}}\frac{1}{%
D_{1}D_{2}D_{3}}\mathcal{~}  \notag \\
&&\times \left\{ \ \left( \mathcal{A}_{1}-\mathcal{V}_{1}\right)
^{B}(x_{123})\left( A_{1}+V_{1}\right) ^{B}(y_{123})\
x_{1}(x_{2}(y_{2}-y_{3})-\bar{y}_{1}y_{2})\right.   \notag \\
&&+\left( \mathcal{A}_{1}+\mathcal{V}_{1}\right) ^{B}(x_{123})\left(
A_{1}-V_{1}\right) ^{B}(y_{123})\ x_{2}(x_{2}-y_{2})(y_{1}-y_{3})  \notag \\
&&\left. +\left( \mathcal{T}_{21}-\mathcal{T}_{41}\right)
^{B}(x_{123})T_{1}^{B}(y_{123})\ 2x_{3}(x_{2}(y_{1}-y_{2})+y_{2}\bar{y}%
_{3})\ \right\} ,  \label{JBE}
\end{eqnarray}%
where $\bar{y}_{i}=1-y_{i}$.  Recall that we only consider twist-4
three-quark operators and neglect the contributions from  twist-4 quark-gluon operators. 

The factor $(1+\delta _{\Lambda B})$ in Eq.(\ref{calGEMB} )  takes into account the correct symmetry  coefficients: 
 the diagrams with two identical fermion lines have the  symmetry coefficient $1/2$. 
 Notice also that  $\Sigma $-baryon DAs are defined as  in Ref.~\cite{Wein:2015oqa}
\begin{equation}
\text{DA}^{\Sigma }\equiv \text{DA}^{\Sigma ^{-}}=-\text{DA}^{\Sigma ^{+}}=\sqrt{2}\text{DA}^{\Sigma ^{0}},
\end{equation}%
which is also taken into account in Eq.(\ref{calGEMB}).

The expression for the amplitude $\mathcal{G}_{E}^{N}$ given in Ref.~\cite{Kivel:2019wjh}
agrees with one given in Eq.(\ref{calGEMB}). The properties of the DAs and the hard kernels
allow one to simplify the expression for the  integrand 
in Eq.(\ref{JBE}) excluding the contributions with $\mathcal{A}_{2},\mathcal{%
V}_{2}$ and $\mathcal{T}_{22}-\mathcal{T}_{42}$.  The  integral$J_{E}^{B}$ is well defined that can be easily seen using the following
observation: {\it all} DAs in the integrand  in   Eq.(\ref{JBE})  have the following
structure
\begin{equation}
\text{DA}(x_{123})=x_{1}x_{2}x_{3}\times \sum_{k_{i}\geq
0}C_{k_{1}k_{2}k_{3}}\text{ }x_{1}^{k_{1}}x_{2}^{k_{2}}x_{3}^{k_{3}}.\text{ }
\end{equation}%
Therefore the singular denominator $(x_{1}x_{2}x_{3}y_{1}y_{2}y_{3})^{-1}$
is  compensated that eliminates a possibility to get the endpoint  singularities.  The resulting integrals for $J_{M,E}^{B}$ can be easily computed numerically
using standard integration packages accessible  in \textit{Wolfram
Mathematica}. 

Let us also mention that the  baryon integrals $J_{M,E}^{B}$ for different baryons  are
equal to each other in the exact $SU(3)$ limit. The formal consideration of this point obviously involves  the identities in
Eqs.(\ref{SU3Phi+}) and (\ref{SU3Phi-}). However these equations do not lead to a simple analytical expressions at the end. 
 In addition the equality of the integrals in the $SU(3)$  limit imposes non-trivial relations between the different  the hard kernels
  in Eqs.(\ref{JBM}) and (\ref{JBE}). It has been explicitly verified  that all such
relations are satisfied, which provides a powerful check of the obtained analytical  expressions.

\section{Phenomenology}

\label{phen}

In the previous section we described the hadronic amplitudes, which are
associated with the three gluon annihilation $J/\psi \rightarrow
3g\rightarrow B\bar{B}$. In order to confront  theoretical predictions
with the  experimental data one has also to take into account the
electromagnetic decay process  $J/\psi \rightarrow \gamma ^{\ast
}\rightarrow B\bar{B}$.  Corresponding amplitudes are described by the
baryon electromagnetic time-like form factors (FFs).  There are strong
indications that these quantities are dominated by long distance dynamics
and therefore can not be accurately computed in pQCD.  In this work we
assume that such contributions provide a sizeable but not very large or
dominant effect and therefore can be ignored at the first step. Therefore
our main task is to study the numerical effect provided by the three-gluon
mechanism to the branching ratio and to the ratio $\gamma ^{B}=|\mathcal{G}%
_{E}^{B}/\mathcal{G}_{M}^{B}|$, which completely defines the angular behaviour
through the parameter $\alpha^g _{B}$%
\begin{equation}
\alpha _{B}^{g}=\left( 1-\frac{4m_{B}^{2}}{M_{\psi }^{2}}\left\vert \frac{%
\mathcal{G}_{E}^{B}}{\mathcal{G}_{M}^{B}}\right\vert ^{2}\right) \left( 1+%
\frac{4m_{B}^{2}}{M_{\psi }^{2}}\left\vert \frac{\mathcal{G}_{E}^{B}}{%
\mathcal{G}_{M}^{B}}\right\vert ^{2}\right) ^{-1},
\end{equation}%
where the superscript $``g"$ indicates that this is pure hadronic
contribution associated with the three-gluon annihilation subprocess.

One more  correction  is described by the combined annihilation $%
J/\psi \rightarrow \gamma ^{\ast }gg\rightarrow B\bar{B}$ . We estimate 
this contribution as a higher order correction and therefore
it will be  excluded from the current analysis.

 For the branching ratio we also use simplified  expression 
\begin{equation}
Br[J/\psi \rightarrow B\bar{B}]=\frac{1}{\Gamma _{J/\psi }}\frac{M_{\psi
}\beta }{12\pi }|\mathcal{G}_{M}^{B}|^{2}\left( 1+\frac{2m_{B}^{2}}{M_{\psi
}^{2}}\gamma _{B}^{2}\right) ,\   
\label{BrR}
\end{equation}%
where $\gamma _{B}=\left\vert \mathcal{G}_{E}^{B}\right\vert /|\mathcal{G}%
_{M}^{B}|$, $\beta =\sqrt{1-4m_{B}^{2}/M_{\psi }^{2}}$ $\ $and the total
widh $\Gamma _{J/\psi }=$ $93$ MeV.

The amplitudes $\mathcal{G}_{M/E}^{B}$ depend on the non-perturbative
parameters, which describe the overlap with initial and final hadrons. \ The
charmonium matrix element is defined in NRQCD as%
\begin{equation}
\left\langle 0\right\vert \chi _{\omega }^{\dag }(0)\gamma ^{\mu }\psi
_{\omega }(0)\left\vert J/\psi (P)\right\rangle =\epsilon _{\psi }^{\mu
}~f_{\psi },  \label{me:fpsi}
\end{equation}%
where the coupling $f_{\psi }$ is related with the quarkonium radial wave
function at the origin 
\begin{equation}
f_{\psi }=\sqrt{2M_{\psi }}\sqrt{\frac{3}{2\pi }}~R_{10}(0).
\end{equation}%
The value $R_{10}(0)$  has been estimated in the various potential models
and in this work we use the estimate obtained for the Buchm\"uller-Tye
potential \cite{Eichten:1995ch} 
\begin{equation}
\left\vert R_{10}(0)\right\vert ^{2}\simeq 0.81\text{GeV}^{3},  \label{R10BT}
\end{equation}%
which implies the charm quark mass to be $m_{c}=1.48\,$GeV.

The baryon and antibaryon matrix elements are defined in terms of DAs as
described above. The model ABO1 is discussed in Ref.~\cite{Anikin:2013aka}, the corresponding
set of the parameters allows one to get reliable description of the
electromagnetic nucleon form factors.  The advantage of this  model is
that corresponding twist-3 and twist-4 DAs provide the unified theoretical
description.  However  this model does not fix the normalisation coupling $%
f_{N}$.  In this work we take  $f_{N}(4$GeV$^2)=4.80\times 10^{-3}$ GeV$^{2}$%
, which is consistent with the sum rule  calculation from Ref.~\cite{Chernyak:1987nv}.  This
value is sufficiently larger than the lattice result $f_{N}^{\text{lat}}(4$GeV$^2)=3.54\times 10^{-3}$ GeV$^{2}$.

The DAs  for other octet baryons are not well known yet. The twist-3
moments have been investigated  within the various frameworks, see  {\textit e.g.}
Refs.~\cite{Chernyak:1987nv, Bolz:1997as}. However these considerations do not include the  higher twist 
DAs.  Recently some of the twist-3 and twist-4 moments have been calculated
on the lattice in Ref.$\,$\cite{Bali:2019ecy}. These results are quite interesting however 
the obtained values for the nucleon DAs are sufficiently different from the
ABO1 model.  In some cases this provides a substantial numerical effect and leads to a
strong disagreement with the experimental data.  For instance, for nucleon the largest numerical effect is related with the
relatively small value of the coupling $f_{N}$ obtained in Ref.$\,$\cite{Bali:2019ecy}. 
Nevertheless we use  the results  from Ref.$\,$\cite{Bali:2019ecy} as a first guess for
baryons $\Lambda ,\Sigma ,\Xi $ and  modify the values of some parameters  in order to
get a more reliable description if the discrepancy with the data are  large.  

The twist-4 DAs depend explicitly from the strange quark mass, see Eqs.(\ref{calVi})-(\ref{calT2i}).  For this mass we take the value $m_s(2\,\text{GeV}^2)=100\,$MeV.  

The  values of different parameters are given in Tab.$\,$\ref{twist34} for twist-3 and twist-4 DAs, respectively.
\begin{table}[th]
\caption{The  parameters, which define the twist-3  and twist-4  models of the baryon DAs (upper and bottom tables, respectively).  All values are given at the scale  $\mu^2=4~$GeV$^2$. The values, which are obtained from the naive $SU(3)$-symmetry are shown by the asterisk. \\}
\centering
\begin{tabular}{|l|l|l|l|l|l|l|l|l|l|}
\hline
$B$ & $f_{B},$ GeV$^{2}$ & $\phi _{10}$ & $\phi _{11}$ & $\phi _{20}$ & $\phi
_{21}$ & $\phi _{22}$ & $f_{\bot }^{B},$ GeV$^{2}$ & $\pi _{10}^{B}$ & $\pi
_{11}^{B}$ \\ \hline
$N$ & $4.80\times 10^{-3}$ & $0.047$ & $0.047$ & $0.069$ & $-0.024$ & $0.15$
& $-$ & $-$ & $-$ \\ \hline
$\Lambda $ & $4.87\times 10^{-3}$ & $0.125$ & $0.050$ & $0$ & $0$ & $0$ & $-$
& $0.044$ & $-$ \\ \hline
$\Sigma $ & $5.31\times 10^{-3}$ & $0.017$ & $0.037$ & $0$ & $0$ & $0$ & $%
5.14\times 10^{-3}$ & $-$ & $-0.017$ \\ \hline
$\Xi $ & $6.11\times 10^{-3}$ & $0.057$ & $-0.0023$ & $0$ & $0$ & $0$ & $%
6.29\times 10^{-3}$ & $-$ & $0.063$ \\ \hline
\end{tabular}
\label{twist34}
\\[4mm]
\begin{tabular}{ | l | l | l | l | l | l | l | }
\hline
$B$ & $\lambda _{1}^{B},$GeV$^{2}$ & $\eta _{10}^{B}$ & $\eta _{11}^{B}$ & $%
\lambda _{\bot }^{B},$GeV$^{2}$ & $\zeta _{10}^{B}$ & $\zeta _{11}^{B}$ \\ 
\hline
$N$ & $-30\times 10^{-3}$ & $-0.037$ & $0.127$ & $-$ & $-$ & $0.127$ \\ 
\hline
$\Lambda $ & $-42\times 10^{-3}$ & $-0.037^{\ast }$ & $0.127^{\ast }$ & $%
-52\times 10^{-3}$ & $-0.037^{\ast }$ & $-$ \\ \hline
$\Sigma $ & $-46\times 10^{-3}$ & $-0.037^{\ast }$ & $0.127^{\ast }$ & $-$ & 
$-$ & $0.127^{\ast }$ \\ \hline
$\Xi $ & $-49\times 10^{-3}$ & $-0.037^{\ast }$ & $0.127^{\ast }$ & $-$ & $-$
& $0.127^{\ast }$ \\ \hline
\end{tabular}
\end{table}
We will refer to these DAs as to unmodified set.

 In order to provide 
reliable estimates for the amplitude $\mathcal{G}_{E}^{B}$ one also needs to estimate
the twist-4 moments, which are not yet been studied.  The  twist-4 moments $\{\eta _{10}^{B},\eta _{11}^{B},\zeta
_{10}^{B},\zeta _{11}^{B}\}$  (except the nucleon case) are not known. Therefore to a first
guess we take for these parameters  the same values  as for the nucleon neglecting 
the $SU(3)$-breaking effects. Further we modify
them in order to improve the description of the data. The corresponding set of DAs
will be referred as modified ones. 

Our  numerical calculations show that the value of $\mathcal{G}_{E}^{B}$ is
 quite sensitive to the parameters $\eta _{11}^{B}$ and $\zeta
_{11}^{B}$.  Therefore these unknown parameters will be 
modified while the computed $\lambda _{1}^{B}$ and $\lambda _{\bot }^{B}$ will
remain unchanged.

The value of the amplitude $\mathcal{G}_{M}^{B}$ is quite
sensitive to the normalisation coupling $f_{B}$.  One finds sufficiently
strong disagreement between the lattice data and the sum rules estimates not
only for nucleon but also for other baryons. The lattice results also indicate about strong $SU(3)$-breaking effects. 
These observations can be interpreted as a presence of uncertainties in the numerical values of $ f_ {B}$.
 Therefore the couplings $f_{B}$ will be also modified in order to improve a qualitative  description.

The DAs parameters depends on the factorisation scale $\mu _{F}$,  more
details about this dependence  can be found in Appendix. The branching ratios strongly
depend on the QCD coupling $\alpha _{s}(\mu _{R}^{2}),$ where  the scale $
\mu _{R}\sim m_{c}$.  In the given analysis we consider  $\mu _{F}=\mu
_{R}=\mu $ and perform the estimates for the two values of the scale  
$\mu^{2}=2m_{c}^{2}$ and $\mu^{2}= 1.5~$GeV$^{2}$ in order to see the value of the corresponding uncertainty.  

The obtained numerical results are presented  in the Tab.$\,$\ref{results}.
\begin{table}[th]
\begin{center}
\caption{The results of the numerical calculations in comparison with the experimental data. 
The obtained values are shown for the  the scale interval  $2m_c^2<\mu^2<1.5~$GeV$^2$. The table includes results for the unmodified and modified DAs. The unmodified   results are shown with the asterisks. \\}
\begin{tabular}{ |c|c|c|c|c|c| } 
\hline
$B$ & 
$Br_{\exp }\times 10^{3}$  &  
$Br\times 10^{3}$ & 
$\gamma _{\exp }^{B}$ & 
$\gamma _{g}^{B}$ &
$\frac{2m_{B}^{2}}{M_{\psi }^{2}}|\gamma _{\exp}^{B}|^{2}$
 \\  \hline
$p$ 
& $2.12(3)$ 
&\multirow{2}{5em}{$0.47-1.43$} 
& $0.83(2)$ 
& \multirow{2}{5em}{$0.66-0.68$} 
& \multirow{2}{2.5em}{ $0.13$} 
\\ 
$n$ & $2.09(2)$ &  & $0.95(6)$ &  & 
 \\  \hline
$\Lambda $ & $1.89(9)$ & $%
\begin{array}{c}
0.27^*-0.81^* \\ 
0.45-1.32
\end{array}
$ & $0.83(4)$ & $
\begin{array}{c}
0.75^*-0.76^* \\ 
0.69-0.69
\end{array}
$ & $0.18$ 
\\ \hline 
$\Sigma ^{0}$ & $1.17(3)$ &  \multirow{2}{7em}{$\left\{ \begin{array}{c} 0.51^*-1.42^*\\ 0.41-1.14\end{array} \right.$ } & $2.11(5)$ 
& \multirow{2}{7em} {$ \left\{ \begin{array}{c}  1.06^*-1.11^* \\  1.68-1.77\end{array} \right.$ }& $1.53$ 
\\ 
$\Sigma ^{+}$ & $1.5(3)$ &  & $2.27(5)$ & 
& $1.31$ 
\\ \hline
$\Xi ^{+}$ & $0.97(8)$ & $
\begin{array}{c}
0.52^*-1.48^* \\ 
0.26-0.74
\end{array}
$ & $0.61(5)$ & $
\begin{array}{c}
0.61^*-0.61^* \\ 
0.59-0.59
\end{array}
$ & $0.22$ \\ \hline
\end{tabular}
\end{center}
\label{results}
\end{table}
In order to estimate the branching ratios we use  in Eq.(\ref{BrR}) the experimental value for the ratio $\gamma^{B}$, which can be easily obtained from the data for $\alpha_B$.  Therefore in this case the only
unknown quantity is the $|\mathcal{G}_{M}^{B}|^{2},$ which is dominated by
the leading power contribution.  The values of the  power suppressed term 
 $\sim |\gamma^{B}_{\exp}|^{2}$  in Eq.(\ref{BrR}) are shown in last column of Tab.$\,$\ref{results}.  
For  nucleon  this term is only about $13\%$, which is not large 
comparing with other expected uncertainties.  For heavier $\Xi$  this  term increases to $22\%$.  However for  $\Sigma $  this term
is  enhanced and its numerical value  becomes even larger than one.   This enhancement  is related with  the  large value  $\gamma ^\Sigma_{ \exp }\sim 2$,  
 which is the direct consequence of the negative polarisation parameter  $\alpha_{\Sigma }< 0$, see Tab.$\,$\ref{dataBB}.  This observation allows one to conclude  that
  for the  $\Sigma $-channel  the  value of the branching ratio  is  not dominated by the amplitude $\mathcal{G}_{M}^{B}$ as for all other baryons.  

The resulting description  of  $\mathcal{G}_{M}^{B}$  shows that for the large scale $\mu^{2}=$ $2m_{c}^{2}$ the obtained branching ratios are about factor $2-3$ below  the data.  For the small scale $\mu ^{2}=1.5$GeV$^{2}$ the agreement with
the data is much better. The main source of the large sensitivity to the
scale dependence in present case  is the value of the $\alpha _{s}(\mu^2)$.  Notice that this  uncertainty cancels in the
ratio $\gamma_g ^{B}$, which becomes quite stable.

The results for the proton-antiproton decay have been discussed in Ref.~\cite{Kivel:2019wjh}. In
Table~\ref{results} we also added the information  for the neutron-antineutron channel. The
experimental value of $\gamma _{n}$ is somewhat larger than $\gamma _{p}$,
but the experimental errors for $\gamma _{n}$ are also larger. Potentially
this relatively small difference can be attributed to the mixing with
electromagnetic FFs.  From the $SU(2)$ symmetry  it follows that $\gamma
_{g}^{p}=\gamma _{g}^{n}$.  Numerically the obtained $\gamma _{g}^{N} $ is
less by $20-30\%$ than $\gamma _{\exp }^{p}$ or $\gamma _{\exp }^{n}$,
respectively. This is rather good result taking into account the underlying
uncertainties. Qualitatively it shows that the  leading-order  contribution
to $\gamma _{g}^{N}$ is sufficiently large and  indicates
that  the factorised contribution  provides  already a reliable description.

The unmodified  DAs in  Tab.$\,$\ref{twist34} provide for $\Lambda $-baryon a relatively small
value of the branching ratio,  which is about factor $2$ smaller  than the
experimental one. At the same time, the obtained value of  $\gamma
_{g}^{\Lambda }$ is  a bit  larger than $\gamma _{\text{exp}
}^{\Lambda }$.  This may indicate  that the amplitude $\mathcal{G}_{M}^{\Lambda }$ is underestimated.  
Therefore in order to make  the description more similar to the nucleon case, one can modify the following one parameter: 
$f_{\Lambda }\rightarrow 5.5\times 10^{-3}$. \footnote{ We assume that the values of the modified  parameters are given at the same normalisation as in Tab.$\,$\ref{twist34}  } 
The larger value $f_{\Lambda }$  increases $\mathcal{G}_{M}^{\Lambda }$ and at the same time this reduces the value of $\mathcal{G}_{E}^{\Lambda}$.  
After that  the obtained results better agrees with the data and qualitatively better overlaps with the description  for the nucleon channel.  
The modified value of $f_{\Lambda }$ is larger than the value obtained from the sum rule
  in Ref.~ \cite{Chernyak:1987nv}: $f_{\Lambda }=4.69\times 10^{-3}$ GeV$^{2}$.  
However the $SU(3)$ violation effect in this case is  rather mild  $(f_{\Lambda}-f_{N})/f_{N}\approx 0.15$.

For the $\Sigma $-decay channel the numerical  results  are different: the unmodified 
DAs provide sufficiently large branching ratio (only for the small scale $\mu $ ) but the  corresponding value of $\gamma _{g}^{\Sigma }$ is about
factor 2 smaller.  Therefore in order to improve the description it is
natural to reduce the value $f_{\Sigma }\rightarrow 4.5\times 10^{-3}$  and
to increase the values $\eta _{11}^{\Sigma }\rightarrow 0.23$ and $\zeta
_{11}^{\Sigma }\rightarrow 0.23$.  This improves the description of $\gamma
_{g}^{\Sigma }$ but also implies sufficiently large $SU(3)$-breaking
corrections for $\Sigma$. The modified values of $\eta
_{11}^{\Sigma }$ and $\zeta _{11}^{\Sigma }$ are about factor two larger then
ones for other baryons. At the same time the modified value $f_{\Sigma }$
is in a good agreement with the  value obtained from the sum rules $f_{\Sigma }=4.65\times 10^{-3}$ \cite{Chernyak:1987nv} . 

In case of the $\Xi$-decay channel the unmodified DAs provide somewhat
larger values for the  branching ratio and for  $\gamma _{g}^{\Xi }$. In
order to reduce these  numbers the following values have been
modified $f_{\Xi }\rightarrow 5.1\times 10^{-3}$,  $f_{\Xi }^{\bot
}\rightarrow 5.29\times 10^{-3}$ and $\eta _{11}^{\Xi }\rightarrow 0.11,\
\zeta _{11}^{\Xi }\rightarrow 0.11$.  The new value $f_{\Xi }$  is more
close to the sum rules result \cite{Chernyak:1987nv}  $f_{\Xi }\rightarrow 4.83\times 10^{-3}$ and
such modification also reduces the effect from the $SU(3)$-breaking
corrections, which is observed for the lattice data. 

\section{Conclusions}
\label{conc}

The decay amplitudes for the process $J/\psi \rightarrow B\bar{B}$  have
been computed within the QCD collinear factorisation framework  for the  octet  baryon states. 
The obtained  results have been used for a qualitative phenomenological analysis. The
primary goal of presented consideration is to estimate the numerical
contributions provided by the factorised amplitudes and to study their
dependence on the models for the baryon light-cone distribution amplitudes.
 In this analysis we do not consider the effect from the mixing with the
baryon electromagnetic decays, which potentially can also provide a numerical impact.  

The obtained results show that the branching fractions for all baryons can
be reasonably described for the relatively low normalisation scale $\mu
^{2}\simeq 1.5\,$GeV$^{2}$ only.  Moreover, the qualitative  description of the branching fractions can be considerably improved if the
values  for  the leading-twist baryon coupling $ f_{B}$, which are obtained from the lattice calculations, 
 are modified. Such modification is very important for the nucleon channel and also allows one to improve the
description for other baryons. Finally the modified set of the couplings $f_{B}$  turns out to be more close to the sum rule estimates obtained in
Ref.~\cite{Chernyak:1987nv}. For this modified set the expected $SU(3)$ breaking effects  are smaller than for the lattice data. 

The obtained ratios $\mathcal{G}_{E}^{B}/\mathcal{G}_{M}^{B}$ describe
the experimental data within the $10\%-30\%$ accuracy, which is quite
reasonable taking into account different theoretical uncertainties. This
indicates that the obtained contributions  provide  sufficiently large
 numerical effect for this observables. The  ratios $\mathcal{G}_{E}^{B}/\mathcal{G}_{M}^{B}$  
 weakly depend on the choice of the QCD renormalisation scale because
  the strong  coupling $\alpha _{s}$ and charm quark mass $m_{c}$ cancel in the ratio. 

The value of the amplitude $\mathcal{G}_{E}^{B}$  is sufficiently large, so that  the ratio  $\mathcal{G}_{E}^{B}/\mathcal{G}_{M}^{B}\sim 0.6-0.8$. 
  The  convolution  integrals for $\mathcal{G}_{E}^{B}$ are  quite sensitive to the shape of the twist-4 DAs, 
 which  can be interpreted  as   three quark state in a $P$-wave. The twist-4  moments $\eta _{11}^{B}$ and $\zeta _{11}^{B}$  for $\Lambda, \ \Sigma$ and $\Xi$ 
 have been  estimated  using $SU(3)$  symmetry and the data.  In the given  analysis  the matrix elements of  twist-4 quark-gluon 
operators are not included  because they contribute to higher order moments, which are not  considered in the used models for twist-4 DAs.   

The experimental data indicate that the value of $|\mathcal{G}_{E}^{\Sigma
}/\mathcal{G}_{M}^{\Sigma }|$ is about factor $2-3$ larger comparing with
other octet baryons. This leads to the interesting observation: the power
suppressed contribution in expression for width, see  Eq.(\ref{BrR}),  is strongly enhanced and provides the very large
numerical effect for the branching ratio. The dynamic origin of this
effect is not clear.  In order to describe this effect within the considered framework  it is necessary  to  assume 
sufficiently large $SU(3)$-breaking corrections.  This implies that  the twist-4 parameters $\eta _{11}^{\Sigma }$ and $\zeta _{11}^{\Sigma }$, 
 are about factor 2 larger comparing with for the nucleon
ones. It remains unclear whether one can explain this enhancement of $|\mathcal{G}_{E}^{\Sigma
}/\mathcal{G}_{M}^{\Sigma }|$  by some intrinsic properties of  the baryon wave
functions or perhaps this effect  may also be related with the hadron dynamics at large distances such as final
state  interactions. 

In conclusion let us  briefly discuss different effects, which can
be important for a more advanced analysis.  
Accounting for interference with electromagnetic amplitudes can improve phenomenological analysis.
  New data obtained for  baryon time-like electromagnetic
form factors allows one  to improve  the  estimate of  this
effect, this work is in progress. 

From the given analysis it also follows that the obtained  description
gives reliable estimates at some relatively low scale only. The scale
ambiguity can be better understood performing the calculation of the next-to-leading
QCD corrections. Such calculation will allow one  to clarify an applicability of the 
collinear factorisation in exclusive charmonia decays.
However the computation of the next-to-leading corrections is quite 
challenging because it involves a big number of the various one-loop diagrams.

 In order to see a size of possible hadronic effects one can  compute the power corrections to the leading-order
amplitude $\mathcal{G}_{M}^{B}$. Such calculation also involves three-quark
twist-4 and twist-5 DAs, which have been already studied in the literature \cite{Braun:2000kw,Anikin:2013aka}.  One can not exclude that the
three-quark twist-4 DAs also provide a large power correction of order $\Lambda
^{2}/m_{c}^{2}$ to the amplitude $\mathcal{G}_{M}^{B}$.  
 If this  is correct then  the description of the decay observables will be considerably improved.

In addition to the  studied decay  mechanism there is one more contribution
associated with the soft-overlap configuration of the final
baryon-antibaryon state. In this case the heavy quark-antiquark pair
annihilates into three hard intermediate gluons, which then create light
quark and antiquark jets.  The hadronisation of the jets into
baryon and antibaryion involves  interactions with the soft quarks and
closely associated with the dynamical  hard-collinear  scale $\mu _{hc}\sim\sqrt{\Lambda m_{c}}$, where $\Lambda $ is a typical hadronic scale.
Taking into account the realistic value of the charm mass $m_{c}$ one
concludes that $\mu _{hc}$ is quite small. Therefore such
soft-overlap matrix element can be considered as non-factorisable in the
light-quark sector.  It is not difficult to show that such contribution is
of the same order in $1/m_{c}$ as the factorisable  collinear one, \textit{see
e.g.} Ref.~\cite{Kivel:2010ns}.  The corresponding hard coefficient function is described
by the two-loop diagrams and also proportional to $\alpha _{s}^{3}$. Unlike the collinear factorisation contribution,
   such  decay amplitude  has nontrivial imaginary phase associated with the cuts
perturbative diagrams.  The hard-collinear  matrix element(s) can be unambiguously defined  within the soft-collinear
effective theory. Such decay mechanism can also provide a
tangible numerical effect and therefore needs to be studied.  This work is
postponed for a future. 

\section{Acknowledgements}
This work is supported  by the Deutsche Forschungsgemeinschaft (DFG, German Research Foundation)
– Project-ID 445769443.

\section{ Appendix}
\label{AppA}
\setcounter{equation}{0}

For a convenience  we provide in this Appendix  additional important details, which complete
the full description of the DAs given in Sec.$\,$\ref{BDA}.  Let us mention
that our notation coincides with  the notation from Ref.~\cite{Anikin:2013aka}.  The twist-3 
DAs are defined in Eq.(\ref{phi3N}) and Eqs.(\ref{phi3Bpm})-(\ref{PiB3}). The
corresponding polynomials $\mathcal{P}_{ik}(x_{123})$ are orthogonal, see {\it e.g.} Ref.\cite{Braun:2008ia}. They  are defined as
\begin{equation}
\mathcal{P}_{10}(x_{i})=21(x_{1}-x_{3}),~~\mathcal{P}%
_{11}(x_{i})=7(x_{1}-2x_{2}+x_{3}),  \label{P1i}
\end{equation}%
\begin{align}
& \mathcal{P}_{20}(x_{i})=\frac{63}{10}\left[
3(x_{1}-x_{3})^{2}-3x_{2}(x_{1}+x_{3})+2x_{2}^{2}\right] ,~ \\
& \mathcal{P}_{21}(x_{i})=\frac{63}{2}(x_{1}-3x_{2}+x_{3})(x_{1}-x_{3}), \\
& \mathcal{P}_{22}(x_{i})=\frac{9}{5}\left[
x_{1}^{2}+9x_{2}(x_{1}+x_{3})-12x_{1}x_{3}-6x_{2}^{2}+x_{3}^{2}\right] .
\label{P2i}
\end{align}
The moments $\phi _{ik}^{B}$ are multiplicatively renormalisable and their
evolution is given by
\begin{equation}
\phi _{ik}^{B}(\mu ^{2})=\phi _{ik}^{B}(\mu _{0}^{2})\left( \frac{\alpha
_{s}(\mu ^{2})}{\alpha _{s}(\mu _{0}^{2})}\right) ^{\gamma _{ik}/\beta
_{0}},\ 
\end{equation}%
where $\beta _{0}=11-\frac{2}{3}n_{f}$ and the anomalous dimensions $\gamma
_{ik}$ read%
\begin{equation}
\gamma _{10}=\frac{20}{9},\ \gamma _{11}=\frac{8}{3},\ \gamma _{20}=\frac{32%
}{9},\ \gamma _{21}=\frac{40}{9},\ \gamma _{22}=\frac{14}{3}.
\end{equation}%

The parameters $\pi _{1i}^{B}$ in Eqs.(\ref{PiL3}) and (\ref{PiB3}) have the same anomalous dimensions as $\phi
_{1i}^{B}$.  The anomalous dimensions for the normalisation couplings 
\begin{equation}
\gamma _{f_{B}}=\gamma _{f_{\bot }^{B}}=\frac{2}{3}.
\end{equation}

The twist-4 DAs, which are defined  in Eqs.(\ref{PhiN4bar}),(\ref{PsiN4bar})
and (\ref{PhiB4pl})-(\ref{PiB4}) include the folowign polynomials

\begin{equation}
\mathcal{R}_{10}(x_{1},x_{2},x_{3})=4\left( x_{1}+x_{2}-3/2x_{3}\right) ,~ 
\mathcal{R}_{11}(x_{1},x_{2},x_{3})=\frac{20}{3}\left(
x_{1}-x_{2}+x_{3}/2\right) .
\end{equation}

The general expressions for the geometrical twist-3 contributions (or
Wandzura-Wilczek part) in the twist-4  DAs can be found  in Ref.\cite{Anikin:2013aka}. 
Here we only provide the explicit formulas for the described models of twist-3 DAs. This gives for the functions in 
Eqs.(\ref{Phi4N}) and (\ref{Psi4N}) 
\begin{align}
\Phi _{4}^{(3)}(x_{123})=& 40x_{1}x_{2}\left( 1-2x_{3}\right)
-20x_{1}x_{2}\sum_{k=0,1}\phi _{1k}\left( 3-\frac{\partial }{\partial x_{3}}%
\right) x_{3}\mathcal{P}_{1k}(x_{123})  \notag \\
& -12x_{1}x_{2}\sum_{k=0,1,2}\phi _{2k}\left( 4-\frac{\partial }{\partial
x_{3}}\right) x_{3}\mathcal{P}_{2k}(x_{123}).  \label{Phi4WW}
\end{align}%
\begin{equation}
\Psi _{4}^{(3)}(x_{123})=40x_{1}x_{3}\left( 1-2x_{2}\right)
-20x_{1}x_{3}\sum_{k=0,1}\phi _{1k}\left( 3-\frac{\partial }{\partial x_{2}}%
\right) x_{2}\mathcal{P}_{1k}(x_{213})
\end{equation}%
\begin{equation}
-12x_{1}x_{3}\sum_{k=0,1,2}\phi _{2k}\left( 4-\frac{\partial }{\partial x_{2}%
}\right) x_{2}\mathcal{P}_{2k}(x_{213}).
\end{equation}%
Notice that the differentiations must be computed with the unmodified
expressions of the polynomials $\mathcal{P}_{nk}(x_{i})$ in Eqs.(\ref{P1i})
and (\ref{P2i}) and only after that one can apply the condition $%
x_{1}+x_{2}+x_{3}=1$.

The  similar contributions in Eqs.(\ref{WWPhi}),(\ref{WWPiB}) and (\ref{WWPiL}%
) are given by%
\begin{equation}
\Phi _{4+}^{B(3)}(x_{123})=40x_{1}x_{2}\left( 1-2x_{3}\right)
-20x_{1}x_{2}\phi _{11}\left( 3-\frac{\partial }{\partial x_{3}}\right) x_{3}%
\mathcal{P}_{11}(x_{123}),
\end{equation}%
\begin{equation}
\Phi _{4-}^{B(3)}(x_{123})=-20x_{1}x_{2}\ \phi _{10}\left( 3-\frac{\partial 
}{\partial x_{3}}\right) x_{3}\mathcal{P}_{10}(x_{123}).
\end{equation}%
\begin{equation}
\Pi _{4}^{B(3)}(x_{123})=40x_{1}x_{2}\left( 1-2x_{3}\right) -20x_{1}x_{2}\pi
_{11}^{B}\left( 3-\frac{\partial }{\partial x_{3}}\right) x_{3}\mathcal{P}%
_{11}(x_{123}).
\end{equation}%
\begin{equation}
\Pi _{4}^{\Lambda (3)}(x_{123})=-20x_{1}x_{2}\ \pi _{10}^{\Lambda }\left( 3-%
\frac{\partial }{\partial x_{3}}\right) x_{3}\mathcal{P}_{10}(x_{123}).
\end{equation}

All the twist-4 moments are multiplicatively renormalisable with the
following anomalous dimensions%
\begin{equation}
\gamma _{\lambda 1}^{B}=\gamma _{\lambda \bot }^{B}=-2,\ \ \gamma _{\eta
10}^{B}=\gamma _{\zeta 10}^{B}=20/9,\ \ \gamma _{\eta 11}^{B}=\gamma _{\zeta
11}^{B}=4.\ 
\end{equation}

\end{document}